# Evaluation the Impact of Library and Information Science Master's Degree (MLIS) on Graduates in Iran

*Asefeh Asemi, PhD.*
*Associate Professor, Head of Department of Knowledge and Information Science,*
*University of Isfahan, Isfahan, Iran, asemi@edu.ui.ac.ir*
*Elham Aghajan, MLIS Std.*
*Department of Knowledge and Information Science, University of Isfahan, Isfahan, Iran*

**Abstract**

**Objective:** The study aimed to examine the effect of MLIS degree on graduates in Iran from different dimensions. The study examined the effects of MLIS on scientific progress, the development of subject expertise, employment, individual characteristics, skills and capabilities, and scientific activities of Iran's graduates.

**Method:** The study was a descriptive-survey and researcher-made questionnaire is used for data collection. The population of the study included all graduates MLIS in Iran that their information was available in, "Iranian Library and Information Studies Alumni Database". 212 persons are selected randomly from 547 persons. After confirming the validity and reliability of questionnaire and collecting the answers, data has been analyzed with SPSS software on both levels of descriptive statistics (frequency tables and graphs of relevant data) and inferential statistics (including single-sample t test to examine the hypothesis, ANOVA test to check differences between variables, Tukey test to compare paired variables, and Pearson correlation test to assess the relationship between two variables).

**Results:** The overall effect average of degree of MLIS on Iranian graduates was equal to 3/25. The findings showed that the average effect of MLIS degree associated with each studied factors on the graduates in the country were: Scientific progress (3/13), development of subject expertise (3/27), employment (3/27), individual characteristics (2/75), skills and capabilities (3/48), scientific activities (3/57).

**Discussion:** Based on the results, the effect of MLIS degree on Iranian graduates was more than moderate. Generally, it can be concluded that MLIS courses at universities in the country, can increase the value of a master's degree of graduates at an acceptable level, but is not perfect; it seems that the authorities should increase their efforts to promote the value of a master's degree in graduates.

## 1. Introduction

Books and libraries have deep bonds with genesis the writing; in the other words invention of writing can be considered the starting point for establishing libraries. Before invention of writing, human was unable to record their experiences as a stable and durable, with writing these knowledge these resources increased gradually, and human needed to a place for protection and maintenance of the resources. The first libraries were established in this circumstances and subsequently, development of libraries lead to changes in the profession of librarianship and advances in library science and there were more necessity of librarian education and training

LIS education in Iran has a history more than fifty years. The first efforts to formal and classic training to librarians return to 1938, short-term training courses run by the Ministry of Culture after this date, and finally thought of establishing MLIS was formed in 1965, In the past decade, 50 departments were founded in, Al-Zahra (former Mottahedin), Iran Medical Sciences, Shahid Chamran (former Jundishapur), Shiraz, and Tabriz universities, and subsequently, the associate, bachelor's, master's, and Ph.D. were Developed (Alimohammadi, 2011, p38). Iran's first graduate library and information science was founded in Tehran University in 1345 and gradually established in other universities.

Today, most undergraduates in LIS in Iran are willing to study in one of sub-disciplines of LIS in master level (academic libraries, school libraries, public libraries, and information), while based on research, despite the old history of LIS graduate programs in Iran, this course isn't empty of difficulties.

Research shows that the expansion of library and information science departments in various universities in Iran and the rapid adoption of these courses, without resolving the previous problems, is one of causes which has made difficult these courses in the country's higher education system. although, the Ministry of Science, Research and Technology and Ministry of Health and Medical Education have devoted extensive efforts to improve training and review headlines in recent years, to form appropriate changes in proportion to world progress in this field with implementing necessary reforms, but it is necessary to investigate feedback of reforms and output of universities and graduates and the value of the degrees they have acquired. With this aim, this study examines the effect of a master's degree in library and information science on graduates in this field in Iran. According to research, the value of a academic degree is assessable in different dimensions. This study tried to examine value and effect of MLIS on graduates in this field with following aspects:

a. Scientific progress;
b. Development of subject expertise;
c. Employment;
d. Individual characteristics;
e. Development of individual skills and capabilities;
f. Scientific activities - Research.

**2. Research Background**

Talk about the quantity and quality of education is as old as the history of human thought. For years scholars attempt to educate librarians and specialists in the field of library and information. Historical background of these trains returns to the second half of the nineteenth century in North America (Horrocks, 1986). Evaluating the output value of these educations is also important with regard to the importance of library and information field. Here are some of the researches which are associated with this article in terms of subject and have been introduced as research background.

Mazinani (1998), in reviewing critical skills needed for academic libraries and specialized information centers in Iran, concluded that the priority of needed skills enumerated by librarians in the profession are: Information technology, library software, professional references, computer applications, and store and retrieve information. Horry (1999) in order to discover the status of research in LIS in Iran and its thematic orientation over time, has analyzed 2490 productions in the form of articles, dissertations, and research projects using survey and bibliometric methods analysis, and their findings indicated that the most scientific contribution is related to the article and then the thesis and research projects respectively. Frarey (1970) in a study entitled "Placements and salaries: the 1969 Plateau" examined attraction condition of 4970 LIS graduates from 1951 until 1961 in Canada and the United States. Their results indicate that 1193 of graduates (81 persons in Canada and 1175 persons in USA) had been attracted to non-librarianship professional and 3517 of them (415 in Canada and 3102 in USA) had been attracted to librarianship professionals. Faculty of information and media studies in University of Western Ontario (2004), evaluated employment status of LIS graduates in 2004 among 44 graduates of this University. The results showed that 42 (95%) of these 44 graduates have been absorbed into the labor market which 75 percent of them were in jobs related to Librarianship. Mayer and Terrill (2005) in their study titled "Academic Librarians' Attitudes about Advanced-Subject Degrees" evaluated the views of academic librarians on the importance of higher education in librarianship, their findings showed among their reasons for tendency to higher education in this field are: Development of research skills, increase credibility, and generally improve their job performance. Whittle and Murdoch - Eaton (2005) studied the skills of university graduates and reached to this conclusion that graduates must possess skills such as the use of information and communication technology, planning and organizational skills such as responsibility, and communication skills such as lecturing, report writing and human relationships. Buckley and colleagues (2009) in a research entitled Doctoral competencies and graduate research education: focus and fit with the knowledge economy?" reviewed a variety of skills and abilities required for PhD and master's students , their aim was to examine whether the PhD degree, can differentiate between graduates of these level with master's and undergraduate students . Results showed that there are deficits despite the cost of education and that PhD qualification is still inadequate. Loo, Mitchell, and Rathbon-Grubb (2010) in their study titled "The value of a doctoral education in academic librarianship: the perceptions of PhD librarians" examined the value of this degree; the study population consisted of university librarians with PhDs in the United States and Canada. There were 300 participants, results showed that librarians with LIS PhD believed this degree had great impact in developing their skills in education and research and employment. In this study with evaluating the capabilities of librarians with PhD degree, also a general framework is presented for developing qualifications of this group.

According to studies, several articles has been conducted in different areas related to feedback of LIS different academic degrees in other countries and from different dimensions but there haven't been conducted any study about effect of MLIS on graduates in Iran, therefore this study examines the value of a MLIS in Iran from six dimension including: scientific progress, development of subject expertise, employment, individual characteristics, development of personal skills and capabilities, Scientific activities-Research.

## 3. Research Theoretical Framework

Training human resources in graduate level in different subfields of LIS and also providing required information is essential to build, equipment and manage libraries and document centers documents. MLIS education program is set with regard to current needs of society and taking advantage of previous experience in library education in Iran and with attempt to address shortcomings in previous programs, Given the rapid development of science and technology and the increasing volume of scientific publications and other media and also diverse needs of community using information, qualified and experienced experts in LIS is necessary for managing libraries and documents centers, and organizing resources for proper utilization of resources and offering services proportionate with needs. In addition, necessity of research in this field of science, and educational needs, necessitates training of qualified and talented professionals who can act in research and education fields. Therefore, the predicted purpose of this level of education program is training manages and researchers.

Many components can be considered and examined in investigating the effect of MLIS degree. According to research backgrounds and components reviewed by previous researchers in determining the value of degrees gained in LIS, Six components have been studied based on figure (1) this section briefly describes these six components.

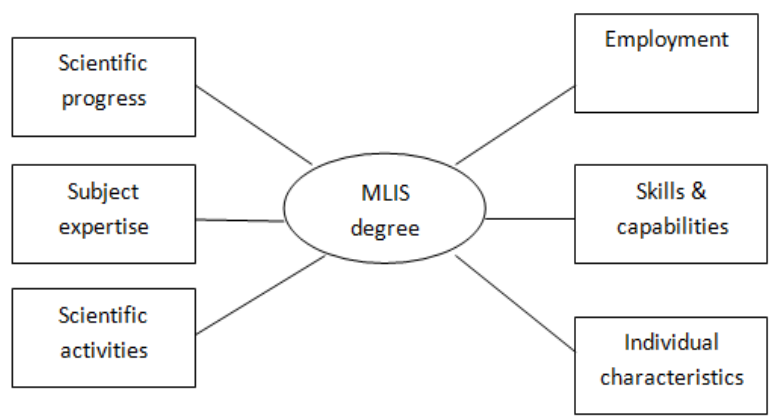

Figure 1. Influenced components after receiving a MLIS degree

### 3-1. Scientific progress

Each scientific discipline need to be supported by national and international publications in various languages to develop and introduce itself to the community. In this regard researchers and professors in every discipline strengthen that discipline with publishing scientific works in form of books, Journal articles, and conference or seminar paper in different languages. For example, the field of LIS was formed first in the United States, and LIS professionals in universities of USA, with publication scientific journals helped to spread this knowledge in the United States and internationally (Norouzi, 2009, pp 28-29). With regard to qualitative and quantitative development in LIS in Iran in the past two decades, which is a function of the development of higher education in Iran, research in this area has evolved widely. Powell, Barker, and Mika (2002) believe that the ability of research to improve services, assistance to planning, and assistance to professional development in LIS is completely unknown. This may be due to suspicion that most of LIS professionals often have to research.

**3-2. Subject expertise**

One of the important missions of higher education is training professional and committed workforce, because skilled and efficient workforce plays a fundamental role in development of any country. In the contemporary world, many of the jobs are based on direct use of information and it. With regard to increasingly specialization of jobs every what more occupations and also becoming expertise necessary for employment micro and detailed, People work as experts in their particular jobs. Librarians and information professionals in created jobs in the field of librarianship and information, as well as other areas, perform their professional activities in new areas of work. These conditions have caused to creation smaller expertise and in return increase in knowledge and skills. On the other hand a rapid development in the areas of employment and advancement of technology has resulted in increased training in LIS. Today in libraries and information centers training is essential for being up to date and effective in performing various tasks. It industry, with multiple products and a variety of employment opportunities, covers a wide range of professional and semi-professional occupations in libraries. Use of technological resources in many libraries, has led specialized and professional works to greater speed and accuracy. Variety of software and applications have provided the employment conditions on the network platform, and these features have caused new expertise is in the library world.

**3-3. Employment**

One of the most important components of educational planning and tutoring for students is attention to their education and professional upbringing. Employments have a very important role in the welfare of people in our society. Humans, gain a major part of their identity from it. Ivez and colleagues (2002) emphasize on this issue that today, education related to new information technology and information systems as one of the most important job skills, Should be an important part of the core curriculum to provide job training in various fields.

UNESCO described the tasks of higher education in employment as it: Higher education should have more diversity in terms of structure such as link between economic conditions and planned courses, pay more attention to public authorities, social skills and to personal development and create regular methods of communication between higher education and world of work. UNESCO have suggested certain communication tools to overcome the problems of supply and demand for educated labor force, include: Employers participation in curriculum development, considering the training curriculum, making flexible the college courses and professional training to part-time workers, part-time work of students in projects that are run by industry, and providing occupation and employment services to help prepare graduates for employment (Harvey, 2000).

It should be noted, despite the high supply of graduates and little demands for employment in libraries , again every year a large number of young people after the entrance examination, are ready to study in universities, thus, they add to this important and complex problem every year.

**3-4. Individual characteristics**

Libraries when have been called as "people's university", which opened their doors to the public and serve to people. Librarian is as the door opener and holder it open. Professional role of librarians in every library and on every levels, is creation a significant interaction between people and information. Thus librarian should posses particular characteristics to accomplish this mission successfully. Specialized and academic skills may not be sufficient for librarians in this way, but before it, they should possess exalted human and moral qualities.

Wilson (1998) believes that the librarians will be successful in performing duties if have characteristics that allow to use them in performing their responsibilities. He argues features for librarians that having these characteristics are necessary for those who want to work in the field of LIS, these characteristics are:

- Flexibility and balance;
- Capability for judgment and non-biasing in decision making;
- Curiosity and risk tolerance;
- Track and stability and endurance in the affairs;
- Management of resources (the resources are: people, time, work processes, tools, and data);
- Pursuit the affairs and Cooperation;
- Knowledge on technical issues of library;
- Problem solving, and offering analytical and intelligent solutions;
- Technical talent.

**3-5. Skills and Capabilities**

In the new era, every day we observe the transformation of higher education and university center. LIS groups are of the most important centers in higher education in the world and Iran that must understand community needs and provide necessary information to meet them. Therefore, given the differences in present and future skills needed by librarians with the skills acquired by traditional librarians, the title of librarians is used in the twenty first century with a different concept from traditional libraries, therefore, changes in the new L IS also leads to changes in the LIS Education. In 21th century's library and information profession in the update training, must first determine which goal the library profession is moving toward, which mission is follow, for determining and explaining education based on them. So understanding the skills needed for the next generation of LIS professionals, should be considered as a key issue for employers and providers of vocational training (Pashutanizadeh, Mansuri, 2008, p144). Librera and colleagues (2004) has emphasized on skills of subject expertise (technical knowledge), communication skills, IT skills, for career success.

Skelton and Abbel (2001) have categorized the skills of librarians have to offer information services on three categories, and each of them can be divided into two sub-categories:

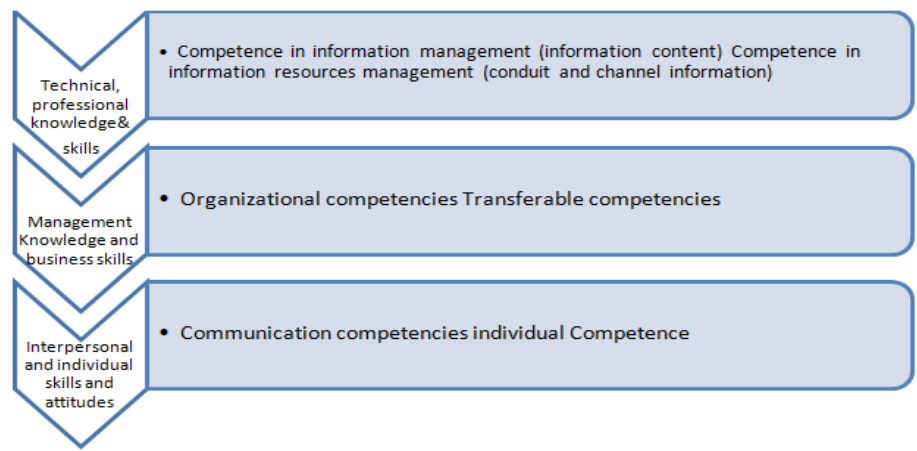

Figure 2. The skills of librarians, libraries (Skelton, & Abell, 2001)

**3-6. Scientific Activities-Researches**

Universities in today's world are known as the mainstream of thinking and thought about the issues which community facing them. Today one of the main categories of interested issues in higher education is "research-based education" and different perceptions of its run manner (Entwistle, 2002). Library is a social institution, which its role and function have been many changes with regard to the global developments, including new technologies and globalization. Librarians as custodians of these developments in the field of libraries will be responsible for new roles and responsibilities. These tasks are raised in different of cultural, research, scientific aspects. Obviously, the ability of research is very effective to improve services, assistance in planning and help to professional development. Booth (2001) posed the theory about "the gap between research and work" in the medical libraries. This theory proceeds to experts fail to apply research findings and the researchers fail to respond to questions related to issues of professionals.

LIS despite having a very strong theoretical foundation is considered as a pragmatic area irrefutably. However, two fundamental issues are raised related to research in the library profession. First, the library professionals rarely consider themselves as active ingredient in the research community. While this group are considered as the subject of many studies and lateral partners in the academic study of and, although the participation of librarians in this issue is limit , often more important that their research rarely is considered serious and important by researchers from academia. The second issue is about dissemination of research results, often the results of research projects won't be disseminated widely to apply in practice. Academic researchers to assess the impact of their research tend to publish results of their research in professional publications published by the experts, but there have been little attempts in publishing their findings (McNicol, 2004).

## 4. Research Methodology

This study was descriptive-survey. This method is applied in the studies which describe one or more quantitative aspect of affairs, and thereby helps the researcher to portray results qualitatively and quantitatively with obtaining facts and comparing them .In this study, researcher-made questionnaire has been used in order to collect data to achieve research purposes and to answer questions of research. The population of this study included all MLIS graduates in Iran that their information was available in, "Iranian Library and Information Studies Alumni Database". 212 persons were selected randomly from 547 persons. Then questionnaire was developed as under- web and sent as e-mail was to them. Data after collecting, were analyzed in descriptive and inferential statistics levels.  frequency tables and graphs of relevant data were used to descriptive analysis, and single-sample t test to examine the hypothesis, ANOVA test to check differences between variables, Tukey test to compare paired variables, and Pearson correlation test were used to inferential statistics**.**

## 5. Research Results

Data obtained from the answers were analyzed, based upon six key components of research, in the form of 81 questions, and in Likert type. Number 3 is considered as the average in this study and data average of each section was compared with the number 3 to determine a MLIS degree whether effect on graduates. The obtained finding in both descriptive and inferential level is as follows:

### 5-1. Data description

Table 1 contains descriptive results including the frequency and frequency percent for "Impact of MLIS degree on graduates in the country" with respect to discussed components.

Table 1. The distribution of respondents views about the "effect of MLIS degree on graduates in the country"

| Component | | Very low | low | Medium | high | Very high |
|---|---|---|---|---|---|---|
| **Scientific Progress** | f | 23 | 36 | 59 | 62 | 22 |
| | % | 11/13 | 17/92 | 27/32 | 30/88 | 10/98 |
| **Subject Progress** | f | 19 | 34 | ۴۸ | 75 | 26 |
| | % | 9/17 | 17/02 | 23/75 | 33/74 | 12/72 |
| **Employment** | f | 45 | 36 | 60 | 46 | 15 |
| | % | 22/43 | 17/96 | 29/36 | 22/72 | 7/5 |
| **Individual Characteristics** | f | 20 | 24 | 34 | 86 | 38 |
| | % | 10/11 | 11/81 | 16/81 | 42/43 | 18/81 |
| **Skills & Capabilities** | f | 13 | 19 | 47 | 85 | 38 |
| | % | 6/36 | 9/65 | 23/21 | 41/89 | 18/89 |
| **Scientific Activities** | f | 19 | 31 | 52 | 77 | 23 |
| | % | 9/22 | 15/05 | 25/71 | 38/10 | 11/32 |
| **Effect of MLIS Degree on Graduates** | f | 5 | 30 | 80 | 83 | 4 |
| | % | 2/5 | 14/9 | 36/9 | 41/1 | 2 |

### 5-2. Data inference

### 5-2-1. Effect of MLIS degree on scientific development of graduates

15 questions were posed in the questionnaires in order to investigate the effect of MLIS degree on scientific development of graduates. According to the findings in table (2) because the probability amount is equal to 0.03 and is smaller than 0.05 therefore (H0) is rejected in significance level of 0.05 and because the mean is equal to 3.13 and is larger than 3 so it is acceptable that effect of MLIS degree on scientific development of graduates is more than average.

Table 2. t-test to evaluate the effect of MLIS degree on scientific development of graduates in this field

| Variable | p | s | mean | df |
|---|---|---|---|---|
| Effect of MLIS Degree on Scientific Development of Graduates in this Field | 0.03 | 2/13 | 3.13 | 201 |

### 5-2-2. Effect of MLIS degree on subject expertise of graduates

13 questions were posed in the questionnaires in order to investigate the effect of MLIS degree on subject expertise of graduates. According to the findings in table (3) because the probability amount is equal to 0.000 and is smaller than 0.05 therefore (H0) is rejected in significance level of 0.05 and because the mean is equal to 3.27 and is larger than 3 so it is acceptable that effect of MLIS degree on subject expertise of graduates is more than average.

Table 3. t-test to evaluate the effect of MLIS degree on subject expertise of graduates in this field

| Variable | P | s | mean | df |
|---|---|---|---|---|
| Effect of MLIS Degree on Subject Expertise of Graduates in this Field | 0.00 | 4.88 | 3.27 | 201 |

### 5-2-3. Effect of MLIS degree on employment of graduates

13 questions were posed in the questionnaires in order to investigate the effect of MLIS degree on employment of graduates. According to the findings in table (4) because the probability amount is equal to 0.000 and is smaller than 0.05 therefore (H0) is rejected in significance level of 0.05 and because the mean is equal to 3.27 and is larger than 3 so it is acceptable that effect of MLIS degree on employment of graduates is more than average.

Table 4. t-test to evaluate the effect of MLIS degree on employment of graduates in this field

| variable | P | s | mean | df |
|---|---|---|---|---|

| Variable | P | s | mean | df |
|---|---|---|---|---|
| Effect of MLIS Degree on Employment of Graduates in this Field | 0.00 | 4.28 | 3.27 | 201 |

### 5-2-4. Effect of MLIS degree on individual characteristics of graduates

Seven questions were posed in the questionnaires in order to investigate the effect of MLIS degree on individual characteristics of graduates. According to the findings in table (5) because the probability amount is equal to 0.000 and is smaller than 0.05 therefore (H0) is rejected in significance level of 0.05 and because the mean is equal to 2.75 and is larger than 3 so it is acceptable that effect of MLIS degree on individual characteristics of graduates is more than average.

Table 5. t-test to evaluate the effect of MLIS degree on individual characteristics of graduates in this field

| Variable | P | s | mean | df |
|---|---|---|---|---|
| Effect of MLIS Degree on Individual Characteristics of Graduates in this Field | 0.00 | -3.61 | 2.75 | 201 |

### 5-2-5. Effect of MLIS degree on skills and capabilities of graduates

8 questions were posed in the questionnaires in order to investigate the effect of MLIS degree on skills and capabilities of graduates. According to the findings in table (6) because the probability amount is equal to 0.000 and is smaller than 0.05 therefore (H0) is rejected in significance level of 0.05 and because the mean is equal to 3.48 and is larger than 3 so it is acceptable that effect of MLIS degree on skills and capabilities of graduates is more than average.

Table 6. t-test to evaluate the effect of MLIS degree on skills and capabilities of graduates in this field

| variable | P | s | mean | df |
|---|---|---|---|---|
| Effect of MLIS Degree on Skills and Capabilities of Graduates in this Field | 0.00 | 6.68 | 3.48 | 201 |

### 5-2-6. Effect of MLIS degree on scientific activities of graduates

25 questions were posed in the questionnaires in order to investigate the effect of MLIS degree on scientific activities of graduates. According to the findings in table (7) because the probability amount is equal to 0.000 and is smaller than 0.05 therefore (H0) is rejected in significance level of 0.05 and because the mean is equal to 3.57 and is larger than 3 so it is acceptable that effect of MLIS degree on scientific activities of graduates is more than average

Table 7. t-test to evaluate the effect of MLIS degree on scientific activities of graduates in this field

| Variable | p | s | mean | df |
|---|---|---|---|---|
| Effect of MLIS Degree on Scientific Activities of Graduates in this Field | 0.00 | 9.70 | 3.57 | 201 |

**5-2-7. The overall objective of the study: determining the effect of MLIS on graduates from various aspects**

Generally according to the findings in table (8) because the probability amount is equal to 0.000 and is smaller than 0.05 therefore (H0) is rejected in significance level of 0.05 and because the mean is equal to 3.24 and is larger than 3 so it is acceptable that effect of MLIS degree on graduates of this field is more than average

Table 8. t-test to evaluate the effect of MLIS degree on graduates in this field

| Variable | p | s | mean | df |
|---|---|---|---|---|
| Effect of MLIS Degree on Graduates in this Field | 0.00 | 4.67 | 3.24 | 201 |

**6. Explanation, Analysis and Conclusion**

A brief comparison between the findings related to the first question with listed studies in the literature shows that findings of Horry (1999) who evaluated the status of research in LIS in Iran and its thematic orientation over time, indicated that the most scientific contribution is related to the article. And also in this study among questions related to scientific progress in the research, productions of domestic articles have the largest contribution. Thus this study confirms the findings of this research. The findings of this research are consisted with findings of Mazinan'i (1998), that review critical skills needed for academic libraries and specialized information centers in Iran, he concluded that the priority of needed skills enumerated by librarians in the profession are: Information technology, library software, professional references, computer applications, and store and retrieve information. On the other hand, comparing findings of Frarey (1970) about the employment of librarians, confirm the findings of this research. His findings such this study suggest that majority of LIS graduates have been attracted to the library profession. In addition, Whittle and Murdoch (2005) studied the skills of university graduates and reached to this conclusion that graduates must possess skills such as the use of information and communication technology, planning and organizational skills such as responsibility, and communication skills such as lecturing, report writing and human relationships. Their findings confirm the results obtained from this research, and in fact the findings also show that the use of ICT skills among graduates is very important. On the other hand, Mayer and Terrill (2005) in their study titled "Attitudes of academic librarians about higher education" evaluated the views of academic librarians on the importance of higher education in librarianship, their findings showed among their reasons for tendency to higher education in this field are Development of research skills, increase credibility, and generally improve their job performance. As you can see the results of this study is consistent with their findings and confirm the effect of MLIS degree on employment and academic credentials and scientific progress and activities of students in PhD and master level. Also, Buckley and colleagues (2009) in a research entitled "the skills of PhD and master" reviewed a variety of skills and abilities required for PhD and master's students and concluded that there are deficits despite the cost of education and that PhD qualification is still inadequate. These results confirm the findings of present study which indicate that these sections are still incomplete and authorities should increase their efforts to promote it. Loo, Mitchell, and Rathbon –Grubb (2010) in their study titled "Evaluation the value of LIS PhD degree in librarians" examined the value of this degree, their results showed that librarians with LIS PhD believed this degree had great impact in developing their skills in education and research and employment. These results confirm the findings of this study, but this fact should be considered also consider that this study had been done on PhD graduates.

As we have seen the effect of master's degree on graduates in this discipline is acceptable but not perfect. The reasons for this, can note to unfamiliarity of this discipline among the people, not updated headlines in graduate courses, the postgraduate level in universities, ignoring the views of graduate students in the curriculum for this level, and the rapid growth of technology. Thus authorities should increase their efforts to promote the value of master's degree in this discipline.

It can be concluded that regardless of the necessity for revisions or modifications in LIS curriculum, this fact must be accepted the that LIS education in the country is faced with challenges, these challenges are not covered to experts, teachers and students, and have provided a lot of concerns about the future of LIS.

If desired statuses of education include indicators such as theoretical foundations, the proper courses, experienced and educated teachers, keep pace with international developments, we must emphasize that this field is now faced with a lot of shortcomings and deficiencies. Deficiencies and shortcomings that likely caused LIS, despite the practical application, cant consolidate its position among other academic disciplines properly . Therefore students in LIS, especially master's students aren't satisfied by their social status, employment status, skills and activities and their reputation, and this makes the MLIS not be perfect.

Hence, the field of LIS as an academic discipline with a brilliant background in the country's higher education system suffers from the problems that may seriously threaten the future of this field. Thus, there is a necessity that requires planners to act to make the necessary changes in its structure over the time. Fortunately, there have done worthy attempts to update the headlines of LIS. Certainly, when LIS can continue to exist in the system of higher education that all teachers, scholars, experts and students in this discipline act in a manner that ,in addition to maintain its philosophy, provide a way to better practical affairs in the IT world.

## 8. References

Alimohammadi, D. (2011). History of LIS in Iran from the view point of a master [Persian]. *Kolliat*, 14 (166), 38-41.

Booth, A. (2001). Research column: turning research priorities into answerable questions. *Health Information and Libraries Journal*, 18 (2): 130-132.

Buckley, F., Brogan, J., Flynn, J., Monks, K., Hogan, T., & Alexopoulos, A. (2009). *Doctoral competencies and graduate research education: focus and fit with the knowledge economy?* [Online]. The Learning, Innovation and Knowledge Research Centre, Dublin City University. Available: http://doras.dcu.ie/2425/1/wp0109.pdf. [22 Jun 2011].

Entwistle, N. J. (2002). Research – based university teaching: what is it and could there be an agreed basis for it? *psychology of education review*, 26 (2), 3-9.

Faculty of Information and Media Studies in University of Western Ontario (2004). *Master of Library and Information Science placement survey*. [Online]. Available:

http://www.fims.uwo.ca/mlis/careers/ placement. [22 Jun. 2012].

Frarey, C. J. (1970). Placements and salaries: the 1969 Plateau. *Library Journal*, 95 (11), 2099-2103.

Harvey, L. (2000). New realities: The relationship between higher education and employment. *Tertiary Education and Management*, 6 (1), 3-17.

Horry, A. (1999). Evaluation, consumption and production manner of scientific information among experts in the country [Persian]. *Etela'resani: negareshha , pajoheshha*, Tehran: Ketabdar.

Horrocks, N. (1986). North American trends in library and information science. *Canadian Library Journal,* 43(5): 293-296.

Ives, B., et al. (2002). What every business student needs to know about information systems. *Communications of the Association for Information Systems*, 9 (1), 467- 477.

Librera, W. L., Eyck, R.T., Dolan, J., Brady, J., & Aviss-Spedding, E. (2004). *New Jersey professional standards for teachers and school leaders*. Newjersy: Department of Education. [Online]. Available: http://www.state.nj.us/education/profdev/profstand/. [22 Jun 2011].

Loo, J., Mitchell, E., & Rathbon-Grubb, S. (2010). *The value of a doctoral education in academic librarianship: the perceptions of PhD librarians*. [Online]. Available: http://sites.google.com/site/phdlibrarians/researches. [22 Jun. 2011].

McNicol, S. (2004). Is research an untapped resource in the library and information profession? *Journal of Librarianship and Information Science*, 36 (3): 119-126.

Mayer, J., & Terrill, L. J. (2005). Academic Librarians' Attitudes about Advanced-Subject Degrees, *College & Research Libraries*, 66 (1), 59-73.

Mazinani, A. (1998). Evaluation the required skills for librarians working in academic libraries and information centers in Iran [Persian]. *Faslnam-e-Ketab,* 9 (1), 44-64.

Norouzi, A. (2009). Evaluation the Scientific productions of LIS professionals in Iran based on existing international articles citation database: Web of Science (1971-2008) [Persian], *Kolliat*, 28-37.

Pashutanizadeh, M., & Mansuri, A. (2008). Library and information profession in the twenty-first century [Persian], *Faslnam-e-Ketab*, 75, 137-156.

Powell, R. R., Baker, L. M., & Mika. J. J. (2002). Library and Information Science Practitioners and Research, *Library and Information science Research*, 24 (1), 49-72.

Skelton, V., & Abell, A. (2001). *Developing skills for the information services workforce in the knowledge economy: a report on the outcomes of eight scenario planning workshops*. London: TFPL.

Wilson, T. C. (1998). *The systems librarian: designing roles, defining skills*. Chicago: American library association publishing.

Whittle, S. R., & Murdoch – Eaton, D. (2005). Curriculum 2000: have changes in sixth form curricula affected students key skills*? Journal of futher and higher education*, 29 (1), 61-71.